# TREBUCHET: Fully Homomorphic Encryption Accelerator for Deep Computation

David Bruce Cousins, Yuriy Polyakov, Ahmad Al Badawi
Duality Technologies
{dcousins, ypolyakov, aalbadawi}@dualitytech.com

Matthew French, Andrew Schmidt, Ajey Jacob, Benedict Reynwar, Kellie Canida, Akhilesh Jaiswal, Clynn Mathew
USC, Information Sciences Institute
{mfrench, aschmidt, ajacob, breynwar, kcanida, akjaiswal, cmathew}@isi.edu

Homer Gamil, Negar Neda, Deepraj Soni, Michail Maniatakos, Brandon Reagen
New York University
{homer.g,nn2231, dss545, michail.maniatakos, bjr5}@nyu.edu

Naifeng Zhang, Franz Franchetti
Carnegie Mellon University
{naifengz, franzf}@cmu.edu

Patrick Brinich, Jeremy Johnson
Drexel University
{pbrinich, jjohnson}@drexel.edu

Patrick Broderick, Mike Franusich
SpiralGen, Inc
{patrick.broderick, mike.franusich}@spiralgen.com

Bo Zhang, Zeming Cheng, Massoud Pedram
University of Southern California
{zhangb, chengz, pedram}@usc.edu

*Abstract*— Secure computation is of critical importance to not only the DoD, but across financial institutions, healthcare, and anywhere personally identifiable information (PII) is accessed. Traditional security techniques require data to be decrypted before performing any computation. When processed on untrusted systems the decrypted data is vulnerable to attacks to extract the sensitive information. To address these vulnerabilities Fully Homomorphic Encryption (FHE) keeps the data encrypted during computation and secures the results, even in these untrusted environments. However, FHE requires a significant amount of computation to perform equivalent unencrypted operations. To be useful, FHE must significantly close the computation gap (within 10x) to make encrypted processing practical.

To accomplish this ambitious goal the TREBUCHET project is leading research and development in FHE processing hardware to accelerate deep computations on encrypted data, as part of the DARPA MTO Data Privacy for Virtual Environments (DPRIVE) program. We accelerate the major secure standardized FHE schemes (BGV, BFV, CKKS, FHEW, etc.) at >=128-bit security while integrating with the open-source PALISADE and OpenFHE libraries currently used in the DoD and in industry. We utilize a novel tile-based chip design with highly parallel ALUs optimized for vectorized 128b modulo arithmetic. The TREBUCHET coprocessor design provides a highly modular, flexible, and extensible FHE accelerator for easy reconfiguration, deployment, integration and application on other hardware form factors, such as System-on-Chip or alternate chip areas.

## I. INTRODUCTION

Current digital infrastructure facilitates secure communication over an insecure communication channel using modern encryption schemes. However, the server needs to decrypt the data before performing any computation, raising concerns about data privacy and security. Fully homomorphic encryption (FHE) is a privacy-preserving computation technique that addresses this problem by performing computation on encrypted data without the need to decrypt the input. Sensitive data is encrypted at the source by its owner, sent to the cloud for secure processing, and the encrypted result sent back to parties approved to decrypt it. At no time is the sensitive data available for decryption by unauthorized parties.

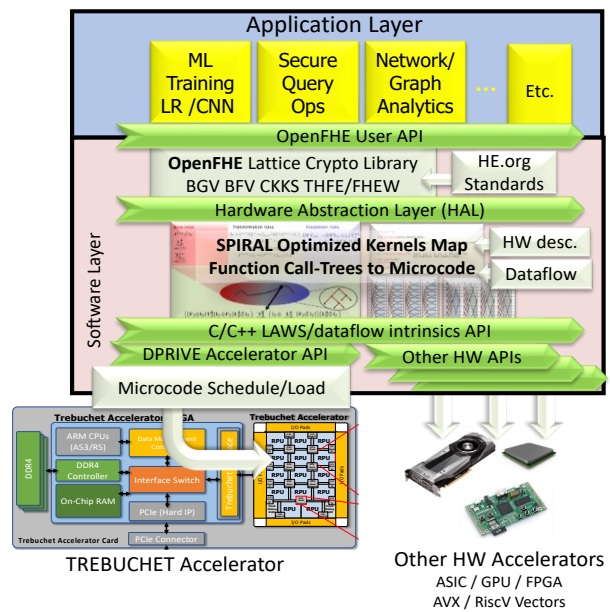

Figure 1 - TREBUCHET Layered System Architecture

FHE is currently used by organizations to analyze shared sensitive data normally restricted by privacy laws. The overhead incurred by FHE is tolerable in these cases because there is no alternative path. However, there are three main barriers to FHE that restrict its wider use. First, conversion of plaintext to ciphertext increases the data size significantly (for example 4B to > 20KB for an integer). This makes data I/O transfer a bottleneck. Second, FHE uses modular arithmetic with a large modulus word size, which is not natively supported by most off-the-shelf hardware, requiring several clock cycles to perform an operation. Third, FHE workloads require extreme amounts of computation vs. their plaintext counterparts (often several orders of magnitude more). These overheads result in high latency, energy consumption, and memory overhead, which limit the applicability of FHE applications.

Several advances in FHE technology have made operation on conventional CPUs more practical, with most schemes supporting encryption of complete vectors, amortizing encrypted math across thousands of elements simultaneously. However, the complexity required for this is large, requiring the use of residue arithmetic to segment the problem into a set of parallel operations, and the use of Number Theoretic Transforms (NTT) to accelerate the convolution of polynomial coefficients during their multiplication. The reader is referred to [1] for detailed discussion of CPU based implementation of FHE operations.

## II. TECHNICAL APPROACH

The fundamental design goal of TREBUCHET is to support 1) a wide array of complex and deep encrypted computing applications, 2) the most important lattice based FHE schemes with 3) a modular design that maps to a wide range of chip sizes with 4) runtime performance orders of magnitude faster than other solutions. We do this by providing basic design blocks and a system stack architecture (Figure 1), that is highly adaptable and extensible. TREBUCHET provides mix-and-match layers for applications, software components and hardware.

Adjacent layers interface using well-defined APIs to encapsulate interactions where our hardware and software interact, allowing each layer to optimize for specific application, scheme, and hardware objectives. We optimized our design through extensive trade space studies between candidate Large Arithmetic Word Size (LAWS) hardware cores, memory, and data flow components. **The Hardware layer** consists of the DPRIVE Accelerator ASIC (DA) on an FHE Processing Board (FPB). The DA is composed of intrinsically modular components easing validation and Formal Verification [2] by reducing the combinatorial explosion of circuit state space. Our approach includes research into novel architectures and their optimizations, on-chip memory systems, crypto data/key reuse/management and optimizing chip I/O bandwidth. We increase performance by maximizing data reuse, limiting communication overhead, trading local vs. global memory and processing near memory to address scheduling and tiling computation onto arrays of LAWS processing elements.

**The Application Layer** consists of C++ implementations of user applications implemented with the OpenFHE[1] API. **The Software Layer** consists of three distinct subsystems to provide high-performance instantiations of all major lattice FHE schemes over a wide array of parameter settings, including all accepted standard security settings [3]. The top layer is the OpenFHE [1] library, which implements all the standard secure FHE schemes in a modular architecture of ring arithmetic required for all major lattice-based encryption schemes. The next lower layer is the SPIRAL NTTX system, which maps high level sequences of OpenFHE lattice crypto function calls into automatically generated software microcode functions (kernels) to program the DA. Finally, a microcode compiler generates firmware instructions for LAWS control sequences.

---
[1] OpenFHE is the successor to the PALISADE library. It shares basic similarities but is engineered to support easier hardware integration using a Hardware Abstraction Layer (HAL). OpenFHE and PALISADE are referred to interchangeably in this document.

**The TREBUCHET DA Architecture**, Figure 2, takes a data-driven approach. A modular parallel, vectorized architecture is used to achieve the highest performance, flexibility, and verification objectives. Ring Processing Units (RPU) are on-chip tiles, that contain multiple ALU lanes for vectorized processing of modulo math with shared vector-data SRAM to buffer ciphertext(s) and keys. Tiles also facilitate memory management by scheduling data to be near computational elements. RPUs are replicated throughout the device, enabling software to minimize data movement and exploit data level parallelism. This architecture provides scalability in the native bit width supported, the number of multipliers and size memory available per tile, and the number of tiles available per chip, enabling system wide optimizations.

## III. HARDWARE LAYER

The Ring Processing Unit (RPU) is designed for general ring processing with high performance by taking advantage of regularity and data parallelism. The RPU utilizes explicitly managed hardware to elide the high costs and complexity of caches, dynamic scheduling logic, and prediction, and task the compiler with scheduling and data movement at compile time. Figure 2 shows an overview of the RPU. Based on the data parallel nature of FHE workloads, parallel vector architectures are highly amenable for meeting the performance needs. We developed an efficient RPU Instruction Set Architecture (ISA) to microcode lattice crypto functions in the RPU. The ISA was co-designed with NTTX and the RPU hardware to address the needs of ring processing while being programmable, as algorithms are still rapidly evolving and to support continued software improvements post fabrication. It has a vector length of 512 elements to maximize work per instruction while providing flexibility to the programmer, as the minimum size ring is typically one to two thousand elements. The ISA includes access to a large, local scratchpad to (double) buffer vector data, multiple vector registers, instructions for register-to-register shuffling of data and native support for large word modular arithmetic. The ISA was designed with simplicity in mind and has only 17 instructions to minimize front-end overheads.

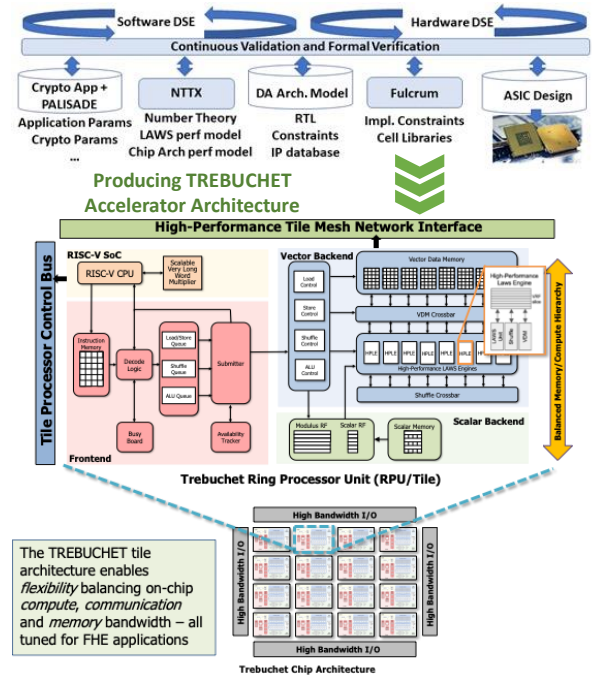

*Figure 2 – TREBUCHET Hardware/Software Co-Design Tool flow and Hardware Architecture Overview*

### A. Frontend

All RPU programs are stored in local instruction memory. When a task is to be executed, a controlling RISC-V core issues a start command to the frontend with an instruction memory pointer to the first instruction of the kernel. To mitigate frontend overheads, in order logic and light-weight dependence tracking is utilized. The frontend fetches and decodes instructions in order. Data hazards are checked using a *busy board*, which is used to describe our light-weight score boarding technique. The busy board is a bit array that tracks all vector registers being used by all inflight instructions. No renaming is supported, and whenever a decoded instruction's register is busy, the entire frontend stalls. The design prioritizes efficiency, the area overheads are negligible, and is highly sensitive to instruction scheduling, which is addressed with the use of SPIRAL.

Once instructions clear all data hazards, they are dispatched to one of three decoupled queues: (1) Load/store Queue, (2) Compute Queue, and (3) Shuffle Queue. Once an instruction is in the queue, it can run in parallel with any other instruction as there are no dependencies. The parallel execution via the decoupled pipelines is key to achieving high performance as it masks much of the data movement time.

### B. RPU Backend

The RPU backend provides the high-performance structures needed for effective ring processing. The major components include three decoupled pipelines for compute via High Performance LAW engines (HPLEs), register-register data shuffling, and Vector Data Memory (VDM). It also includes a scalar memory to house the constants needed by HE.

**1) High Performance LAW Engine (HPLE):** The HPLE is the computational unit in the RPU. Each has a LAWS engine and is partition of the Vector Register File (VRF), or a VRF slice. The LAWS Engine contains a modular multiplier, a modular adder, a modular subtractor, and two comparator units. NTT/iNTT is a key kernel in Ring-LWE, and the HPLES support native butterfly computation via a butterfly instruction.

Each CI command interacts with VRF slice and LAWS Engine to perform three tasks: read data from the VRF slice to the LAWS engine, start the computation in LAWS Engine, and store the output to the VRF slice. Here we use 128b to meet the needs of HE precision. The RPU allocates multiple HPLEs as lanes in classic vector designs.

In each HPLE, the LAWS Engine is connected to the VRF slice. VRF slice is a part of the VRF that is divided among HPLEs. According to our *RPU* ISA, VRF has 64 vector registers with 512 elements. Each slice has $64 \times 512/(num\_HPLEs)$ elements. If we store each register of VRF in different memory, it requires small and efficient memory. To increase area efficiency, we stack four registers in one memory. The four registers in one memory cannot be accessed simultaneously, and SPIRAL handles special scheduling and data placement in the VRF. Hence, a VRF slice has 16 single-port memory with $(512 \times 4)/(num\_HPLEs) words$. A VRF slice interacts with HPLEs, VBAR, and SBAR. To support these parallel connections, each VRF slice has ten ports; five ports (three read ports and two write ports) for HPLEs, three ports (two read ports and one write port) for SBAR, and two ports (one read port and one write port) for VBAR. For computation, each VRF slice sends the data from input registers to corresponding HPLEs. HPLEs performs the computation. Once HPLE outputs the result, VRF slice stores it back to output registers.

**2) Shuffle Crossbar (SBAR):** The SBAR transfers the data across VRF registers, facilitating efficient implementations of complex access patterns to maximize NTT efficiency by allowing register-register data shuffle. With the SBAR, vectors can be broken up in the VRF, saving round trips through the VDM to restructure data in ISA referenced vectors. The SBAR supports all four modes of shuffle transfer.

**3) Vector Data Memory (VDM):** We instantiate an RPU with a 4MiB Vector Data Memory (VDM). Here we find 4MiB is sufficient capacity to double buffer off-chip data loading with the execution of a kernel. The VDM can be up to 32MiB if more space is needed and can be banked to increase bandwidth. The large word size and capacity of the VDM necessitates the use of large SRAM macros that tend to run at low frequency and are currently the bottleneck in a single clock domain design.

**4) Vector Crossbar (VBAR):** The VBAR transfers the data between VDM and VRF slice of HPLEs. It supports four modes of data transfer. When an HPLE reads or write data from different VDM banks, VBAR transfers the data in parallel. In practice, we find striding data across banks resolves nearly all bank collisions. We designed a parameterized VBAR to support any number of banks and HPLEs.

**5) Scalar Backend:** A Scalar Data Memory (SDM) and Scalar Register File (SRF) are included to handle the many constants needed in RLWE processing. The SDM is 32KB and uses 128b words, which it loads into the SRF. The SRF sends values to HPLEs when the RPU executes scalar instruction. To add flexibility of operations, a Modulus Register File (MRF) is part of the backend. The MRF enables modulus changing at the instruction granularity, enabling processing different sets of data simultaneously. SRF and MRF data is directly transferred to HPLEs. To add flexibility of programming, we include an Address Register File (ARF) for indirect memory access.

### C. Implementation

The Trebuchet design requires rapid Design Space Exploration (DSE) to evaluate the optimal selection of topologies and components as circuit-level optimizations for FHE are discovered. To facilitate this, we developed the Fulcrum platform, Figure 3, to rapidly explore and implement different architectures. Fulcrum consists of three core features: *Architecture Modeling, Accelerator Generation,* and *Physical Design*. The Architecture Modeling provides a DSE interface to allow designers to rapidly select and explore FHE design parameters specified by the cryptography and algorithms team. The Accelerator Generator interface enables different system architectures to be assembled based on processing, memory allocation, and interconnection parameters. The Physical Design stage takes the design through the standard ASIC tool flow, where further DSE over modern ASIC EDA tool optimizations and different physical constraints (clock rates, layout etc.) can be explored. Fulcrum was used to assess ~1,000 multiplier, interconnect and memory designs that were collected during Phase 1 to generate Pareto Optimal architectures.

Fulcrum was used to generate and test all low-level components, as well as fully integrated RPU tiles with fast simulation test benches in Python leveraging CocoTB and full hardware emulation on a Palladium system with complex kernels and test vectors. The Fulcrum EDA tool flow is highly modular and given different implementation constraints (i.e. SWAP) can generate different designs including FPGA implementations.

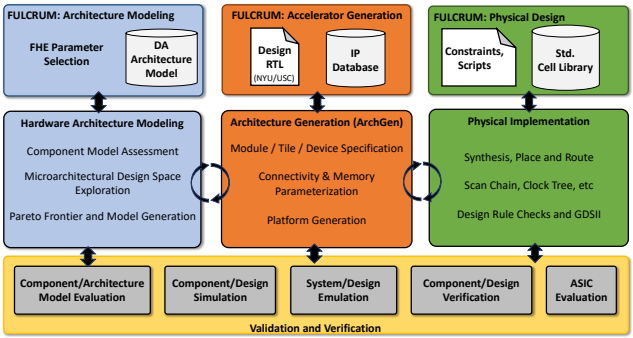

Figure 3 Fulcrum FHE DSE EDA Tool Flow

### D. Component Optimizations

In moving from traditional 32- and 64-bit data word architectures to 128-bit words, critical innovations were required in the modulo multipliers and on-chip data movement structures (crossbars) to address exponential area scaling. We explored over 390 modulo ALU designs, Figure 4, with the selected ALU being a pipelined Barrett modulo multiplier that leveraged optimizations which eliminated a half-size multiplication during computation of the quotient and reduced the cost of calculating the product by only calculating the on the least significant N bits [4]. These optimizations enabled sub-quadratic complexity growth of the multiplier. For crossbar optimizations, we depopulated unused configurations and hierarchical multiplexers, allowing use to achieve an 11.5x area savings, and a design that is 128 to 8,192 times faster than conventional bus based designs used in FHE accelerators.

## IV. SOFTWARE LAYER

Several popular, high quality, open-source libraries exist for implementing systems based on FHE. SEAL from Microsoft Research [5], HElib from IBM research [6], PALISADE [7] and its successor OpenFHE[1] developed by multiple authors including those on this paper from Duality Technologies. We chose PALISADE for our initial work, moving to OpenFHE upon its release. OpenFHE supports all the mentioned schemes, is implemented in C++ and is heavily optimized for vector operations, using residue arithmetic to reduce large bitsize arithmetic into smaller conventional machine words, including 128-bit arithmetic support. The OpenFHE library follows the Homomorphic Encryption Standard by meeting proper bit-security thresholds, given via (semi-)automated parameters that can be set by the user to control what security and performance they prefer to target.

We extended SPIRAL [8] to support NTT and batch NTTs. Mirroring the structure of FFTW and FFTX, the NTTX package offers FFTW-style C/C++ API in line with FFTXstyle code generation, powered by SPIRAL in the backend. Illustrated by Figure 5, NTTX API leverages SPIRAL's capability of delayed execution and just-in-time code generation to implement an inspector/executor paradigm for OpenFHE.

To support general radix NTTs, large vector instructions and simple parallelism in SPIRAL, we added both the KornLambiotte FFT algorithm [8] and the Pease FFT algorithm [10] as breakdown rules to SPIRAL. Using SPIRAL's Operator Language (OL), NTTs of size $r^k$ are represented as

$$NTT_{r^k} = R_r^{r^k}\left(\prod_{i=0}^{k-1} L_{r^{k-1}}^{r^k} D_i^{r^k}\left(NTT_r \otimes I_{r^{k-1}}\right)\right)$$

To execute the generated NTT code, NTTX allows various data types for long vectors, provides different schemes of register allocation (e.g., greedy, naive round robin), and has the infrastructure for verification (e.g., functional simulator) and low-level optimizations (e.g., instruction scheduler).

```
// SPIRAL generated NTT Code for TILE vector architecture
#include <tile.h> void
_ntt1024x512_b1() {
    enter(OP_DEFAULT); _vload_512x128i(REG_V60, REG_A1, 0);
    _vload_512x128i(REG_V20, REG_A1, 8192);
    _vbroadcast_512x128i(REG_V19, REG_A3, 1, 1);
    _vimulmod_512x128i(REG_V59, REG_V20, REG_V19, REG_M1);
    _vaddmod_512x128i(REG_V58, REG_V60, REG_V59, REG_M1);
    _vsubmod_512x128i(REG_V57, REG_V60, REG_V59, REG_M1);
_vunpacklo_512x128i(REG_V56, REG_V58, REG_V57); ...
    _vstores_512x128i(REG_A2, 16, REG_V21, 2); leave(OP_DEFAULT);
}
```

Listing 1: SPIRAL-generated radix-2 1,024-point NTT code using shuffle instructions.

We generated forward and inverse vectorized radix-2 NTTs with sizes from 1,024 to 65,536 and verified their cor-rectness with OpenFHE generated data. Listing 1 shows the radix-2 1,024-point NTT code generated by SPIRAL. To address the interoperation of NTTX to the RPU hardware, we translate NTTX calls into our RPU ISA discussed above – this is straightforward as there is usually a one-to-one correspondence between the two (as they were co-designed with each other in mind).

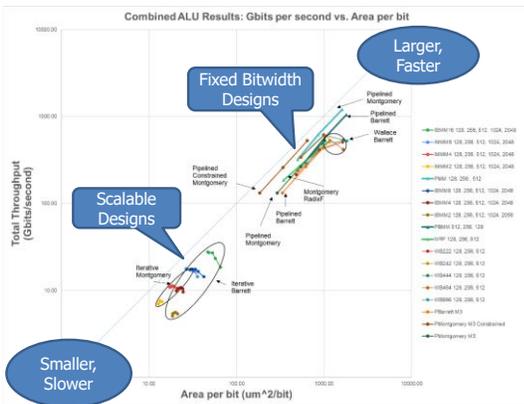

Figure 4 Trebuchet LAWS ALU Design Space

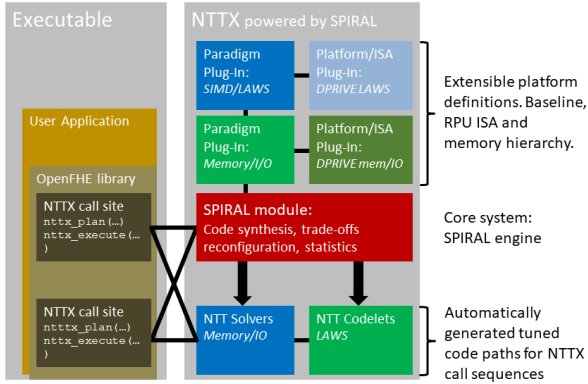

*Figure 5 OpenFHE and SPIRAL NTTX Interaction*

## V. RESULTS

The goals of the project are to maximize the processing within the largest available footprint allowable without waivers on a GF12LP Multi-project Wafer (MPW), or 150mm$^2$. The first phase of this project documented in this paper focused on development of the LAWS computations and the ring processing unit. The second phase of this project, which is currently underway, scales the design to the full device, integrates I/O and other peripherals, and fabricates the full device. Therefore, for the results in this paper, we characterize the performance of the RPU using AFRL's Palladium emulation system, and project full device performance based on different device level topologies, memory, and I/O interfaces.

We emulated a single RPU on Palladium, with an ALU Size of 113K gates, a Lane size of 137K gates and a total RPU Size of 11.83M gates. The ALU operations include 128-bit Scalar and Vector operations across Addition, Subtraction, Comparisons, Modular Multiplications. For kernel operations these were evaluated across an increasing kernel size of 1024, 16K, and finally 64K points for Switch Modulus, Fast Basis Extension, and RingMul, RingAdd, NTT (all verified in the same run). The resulting Palladium trace verifies the number of clock cycles needed to execute the kernel.

### A. Application Mapping Considerations

We developed an approach to characterize the workload for a Logistic Regression (LR) training application. We used an LR algorithm [9] that was friendly to our BGV Fixed Point Packed Encoding [10] and implemented it in the PALISADE library to verify correct operation. To estimate runtime performance, we accumulated basic vector operation counts, developing a model of the required software operations and their associated complexity numbers in the form of accumulated ringAdd, ringMult and NTT/iNTT counts. We modeled functional bootstrapping for a sigmoid approximation, sign() determination and rescaling of the two vectors that are updated in the logistic regression training loop.

We selected parameters for the BGV implementation that provided sufficient accuracy for the logistic regression training, resulting in a ring size of $2^{16}$. These parameters allow us to pack 64 SIMD slots per ciphertext. The parameters were chosen such that we have 7bits of fractional precision and a maximum data range of 127bits. We scaled our input data so that no overflows occur during the Logistic Regression training.

Our Logistic Regression Training required the input Ciphertexts (CT) to have a multiplicative depth of 2, or three towers (each tower is based on a LAWS residue of a large modulus, also called a "ring"), except during bootstrapping where they are expanded to ~30 towers.

We determined that the total number of bootstrap operations per logistic regression iteration loop is 32 for sigmoid (compare operation), 16 for rescaling the input to sigmoid + one for sign() and two for rescaling the weight vector (b or beta) and the residual vector (res) for use in the next iteration, for a total of 33 sign/compare bootstraps and 18 rescaling bootstraps for a total of 51 functional bootstraps per training epoch. Bootstrapping is very compute intensive and composed primarily of Keyswitching and modulus switching operations. We derived approximate runtime estimates based on modeling the number of these two functions for each of the forms of functional bootstrapping required and converting those to base ring operations.

### B. Chip I/O Modeling

Trebuchet's I/O bandwidth can scale to available resources, and considers Low Voltage Differential Signaling (LVDS) at Single Data Rate (SDR) and Double Data Rate (DDR) for between 62.5 GB/s and 250 GB/s bandwidth to the chip. We have done backend layout of RPU tiles and have budgets for the total chip area estimate using 14 tiles and I/O controllers. We also consider HBM and HBM2 memory interfaces, which can provide 288 GB/s, and 460.8 GB/s respectively. We developed a model to account for the extra time required for data to load/unload from the chip depending on the interface technology. Our internal tile mesh interconnect is capable of scaling to any of the above data rates so it will not limit the data rate to an RPU.

The current RPU configuration has 4 MB of Vector Data Memory, which is enough storage for four 64k * 128-bit towers. There is enough storage in an RPU to buffer two vectors while operating on the main data and twiddle memory during NTT operation. We use this in our scheduling to hide the data loading for NTTs, thus our NTT timing is unaffected by data motion.

Additional overhead to load data into the chip depends on the application and involves complex analysis. We instead use a conservative simple data model with some operations throttled by chip I/O data rate. We determined that during Encrypted ciphertext multiplication, RPU memory would not be sufficient to hold all data vectors for a single tower, requiring a reload of some vector data on-chip, stalling our ALUs while a new operand is loaded. Thus, we derated the timing of *all* ring multiply operations to the time it takes to load or store one vector of tower data for the various I/O interfaces. This should be taken as a conservative worst-case number.

*C. Performance Results*

Our current best values from Palladium emulation of 64k point RA and RM are RA = RM = 1.73 uSec. Execution time for NTT is a bit more complex. We have two versions of the RPU that currently have different NTT execution times. The two candidate RPU Tile configurations, RPU version 1 with one vector register per data memory unit and RPM Version 2 with four vector registers per memory unit. The former is bigger and easier for SPIRAL to schedule, the latter is smaller and more difficult to schedule. Our fully functional and Palladium validated 64k NTT running on RPU Tile 1 will run in 18.3k clock cycles or 9.15 uSec. We also have a version of NTT currently running at 24.6k clock cycles or 12.3 uSec. RPU Model 1 can fit 10 tiles on a chip, while Model 2 can fit 14 on a chip. We believe that with additional improvements in the NTT with SPIRAL in phase 2 we should be able to schedule the NTT to reach an ideal limit of 16k clock cycles or 8 uSec running on the smaller RPU Tile Model 2.

Table 1 shows the resulting run times for the three RPU configurations, across the four data models (ideal, LVDS DDR 2k pins, single HBM2 and two HBM2) allocating the computational load evenly across all RPUs. The number of RPU tiles is reduced by one in the dual HBM2 case to make room on the chip. We show the three Level 0 operations as well as the timing for single sign and rescale functional bootstraps. We also show a derating factor as a % slow-down for a given I/O model relative to the ideal model. It is part of our phase 2 tasking to better balance the compute and I/O.

*Table 1-Timing summary for one Logistic Regression Training Iteration in seconds.*

| Ring Processing Unit (Tile) Configuration | I/O data model | Log Reg iteration time no BS | %Derate from ideal data model | Log Reg iteration time w/BS | %Derate from ideal data model |
|---|---|---|---|---|---|
| RPU Model 1 | Ideal | 0.110 | | 3.650 | |
| RPU Model 2 | | 0.091 | | 2.912 | |
| RPU Phase2 | | 0.074 | | 2.495 | |
| RPU Model 1 | LVDS2k | 0.153 | 39% | 5.370 | 47% |
| RPU Model 2 | | 0.122 | 34% | 4.141 | 42% |
| RPU Phase2 | | 0.105 | 42% | 3.724 | 49% |
| RPU Model 1 | 1xHBM2 | 0.125 | 14% | 4.233 | 16% |
| RPU Model 2 | | 0.102 | 12% | 3.329 | 14% |
| RPU Phase2 | | 0.084 | 14% | 2.912 | 17% |
| RPU Model 1 | 2xHBM2 | 0.122 | 11% | 4.055 | 11% |
| RPU Model 2 | | 0.098 | 8% | 3.136 | 8% |
| RPU Phase2 | | 0.080 | 8% | 2.687 | 8% |

One key take away is that when analyzed with a full application load, the RPU Model 2 performs better than the RPU Model 1, because while it is slower for NTT, it is smaller and there are more tiles of the former than the latter. So overall runtime is better for the second model.

It is important to compare the latency of these operations vs a single CPU running PALISADE with 128bit arithmetic[2]. Table 2 shows the speedup with one RPU, a full chiplet of 14 RPU tiles, and a transition ready configuration of four chiplets on a multi-chip module.

*Table 2 - Comparison of core kernel latency vs. single CPU running PALISADE 128-bit software*

| Kernel | CPU latency (us) | RPU latency (us) | Single RPU Speedup | Single chiplet Speedup (14 RPU tiles) | Quad Chiplet Speedup (4 chiplets) |
|---|---|---|---|---|---|
| ringAdd | 333 | 1.73 | 192x | 2694x | 10779x |
| ringMult | 1,240 | 1.73 | 716x | 10034x | 40138x |
| NTT | 7,807 | 12.3 | 634x | 8886x | 35544x |

## VI. CONCLUSIONS

Our results for the first phase of the DPRIVE project focused on the computational aspects of a modular and flexible design of a tiled FHE coprocessor, demonstrating large speedups (35,500x) over conventional CPU approaches. The next phase will focus on the impact of data marshalling and chip I/O bandwidth for the overall design under the offered load of larger ML applications such as Convolutional Neural Networks. In addition, we will focus on using the CKKS approximate number scheme, which provides more efficient ciphertext packing and greatly reduces the number of bootstrap operations.

---

[2] Using the PALISADE benchmark bin/benchmark/poly-128-benchmark-64k set for a single core on a Dell Precision-3630-Tower: 4700 MHz CPU. CPU Caches: L1 Data 32 KiB, L1 Instruction 32 KiB, L2 Unified 256 KiB, L3 Unified 12288 KiB.